\documentclass[superscriptaddress,preprint,amsmath,amssymb,aps,prl]{revtex4-1}
\pdfoutput=1
\usepackage{times}
\usepackage{graphicx}
\usepackage{subfigure}
\usepackage{epstopdf}

\begin{document}

\title{A Fundamental Limit on Propagation Loss of Waveguides with Subwavelength Mode Size}

\author{Amir Arbabi}
\email{Corresponding authors AA: amir@caltech.edu}
\affiliation{Department of Electrical and Computer Engineering, University of Waterloo, 200 University Avenue West, Waterloo, Ontario, Canada N2L 3G1}
\affiliation{T. J. Watson Laboratory of Applied Physics, California Institute of Technology, 1200 E California Blvd., Pasadena, CA 91125, USA}

\author{Ehsan Arbabi}
\affiliation{T. J. Watson Laboratory of Applied Physics, California Institute of Technology, 1200 E California Blvd., Pasadena, CA 91125, USA}
  
\author{Safieddin Safavi-Naeini}
\affiliation{Department of Electrical and Computer Engineering, University of Waterloo, 200 University Avenue West, Waterloo, Ontario, Canada N2L 3G1}

\begin{abstract}
A fundamental trade-off relation between the cross sectional confinement and propagation length of an arbitrary mode of a general waveguide is presented. This limit is a generalization of the well-known diffraction limit for guided modes. The results provide a lower bound on propagation loss of plasmonic waveguides which are attractive for their deep subwavelength mode dimensions. We also introduce a material loss merit factor which sets a criterion for comparing different plasmonic materials for achieving the best trade-off between confinement and loss.
\end{abstract}

\maketitle

\section{Introduction}
\label{sec:introduction}
In recent years, with the advent of the emerging field of surface
plasmon photonics or ``Plasmonics" \cite{Ozbay2006, Barnes2003}, subwavelength waveguides at optical frequencies have
attracted much attention. Different types of waveguides have been
proposed and investigated for their ability to guide optical waves with subwavelength cross sectional mode dimensions \cite{Bozhevolnyi2006,Zia2004,Dionne2006,Baida2006,Gramotnev2004,
Oulton2008, Pile2005, Oulton2009, Rybczynski2007}. Based on these waveguides, several passive and active devices and elements such as bends, interferometers, filters, 
resonators, and lasers have been demonstrated~\cite{Veronis2005,
Hosseini2007, Boltasseva2005, Pile2005_2}. Although waveguides with
subwavelength mode dimensions such as coaxial and microstrip lines
are widely used at millimeter wave, microwave and lower frequencies without a
significant loss, at higher frequencies metal absorption loss has
been an obstacle for plasmonic waveguides to find
practical applications. In some of the studies, it has been noted that there is a trade-off between the confinement of a waveguide mode and its attenuation constant \cite{Chen2006, Oulton2008}. Here, we show that regardless of the waveguide shape and design, for a desired mode size, there is a fundamental lower bound on its attenuation constant. 

The existence of such a limit is a generalization of the diffraction limit which imposes a lower bound on confinement of free propagating waves. Therefore, we use a similar approach to the one used for deriving the diffraction limit.  The fields are considered in spectral domain and the waveguide is replaced with a equivalent current distribution. This is this current distribution which allows for confinement beyond the diffraction limit, and since it is supported by the lossy waveguide material, it causes the propagation loss. We show that for achieving tighter confinement, larger equivalent currents are required which cause higher propagation loss.

\section{Problem formulation}
\label{sec:Problem formulation}
\begin{figure}[t]
\centerline{\includegraphics[width=0.4\columnwidth]{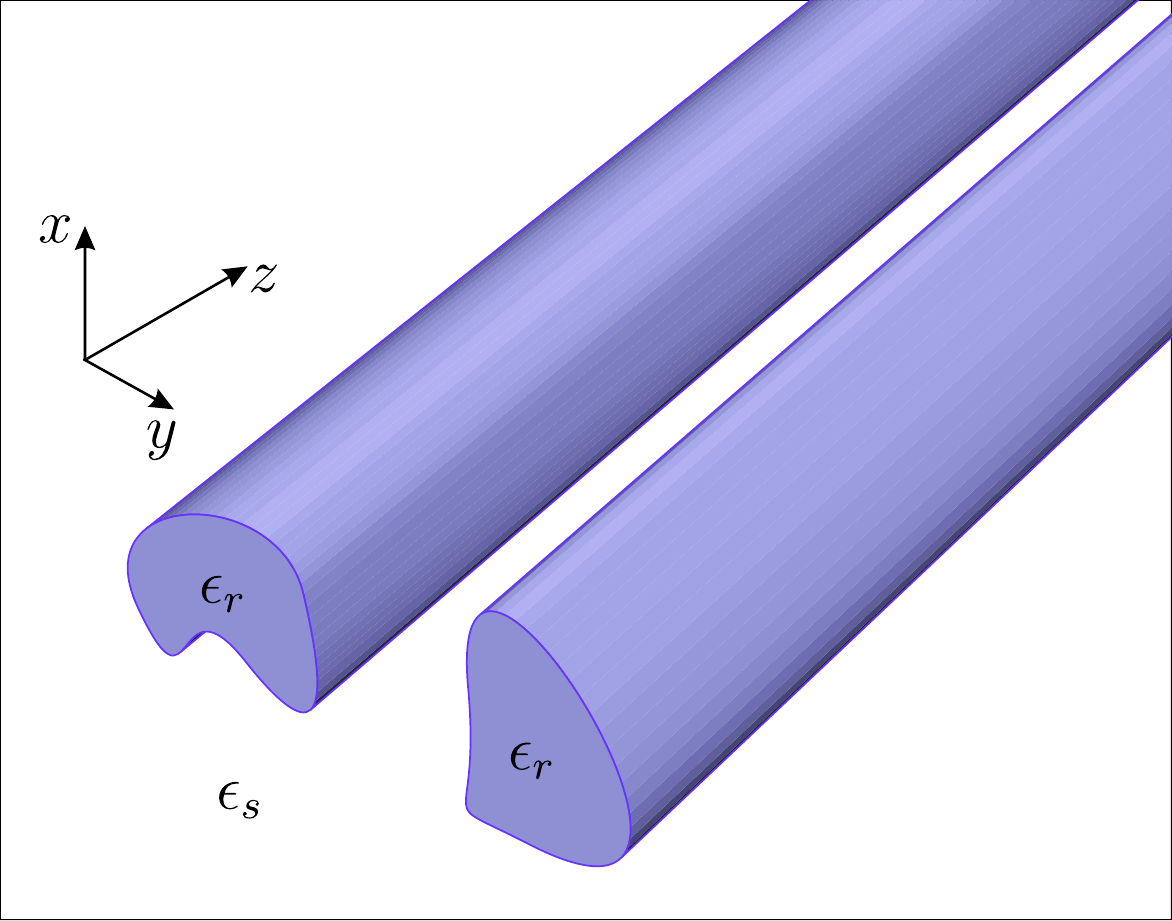}}
\caption{Schematic illustration of a general waveguide composed of lossy dielectrics with relative permittivity of $\epsilon_r$ surrounded by lossless material with permittivity of $\epsilon_s$. $z$ axis of the coordinate system is parallel to the waveguide axis.}
\label{fig:waveguide}
\end{figure}
We consider a general waveguide made of a non-magnetic material with permittivity of
$\epsilon_r=\epsilon_r'-j\epsilon_r'' $ surrounded by a lossless
material with permittivity of $\epsilon_s$  (as shown schematically in Fig. \ref{fig:waveguide}). The surrounding material may fill the entire space around the waveguide or only the space near the waveguide where the fields have
significant values. We assume that the waveguide has no
variations along its axis, and the coordinate system is
chosen in a way that its $z$ axis is aligned with the waveguide axis.
We also assume time dependency of $\mathrm{e}^{j\omega t}$ and 
 $z$ variation of $\mathrm{e}^{-j\gamma z}$,
  $\gamma=\beta-j\alpha$,  where $\beta$ and $\alpha$ are phase and attenuation constants, respectively. The electric and magnetic fields of the waveguide mode satisfy the Maxwell's equations in the entire space
\begin{subequations}
\begin{eqnarray}
&\nabla\times\mathbf{E}=-j\omega\mu_0\mathbf{H}, \label{eq:firstMaxwell}\\
&\nabla\times\mathbf{H}=j\omega\epsilon_0\epsilon_s\mathbf{E}+\mathbf{J},
\label{eq:secondMaxwell}\\
&\nabla\cdot\mathbf{E}=\frac{j}{\omega\epsilon_0\epsilon_s}
\nabla\cdot\mathbf{J},\\
&\nabla\cdot\mathbf{H}=0,
\end{eqnarray}
\end{subequations}
with $\mathbf{J}$ representing the equivalent current density defined as
\begin{equation}\label{eq:J}
\mathbf{J}=j\omega\epsilon_0(\epsilon-\epsilon_s)\mathbf{E}.
\end{equation}
$\epsilon$ is spatially varying relative permittivity describing the waveguide and its surrounding media. Therefore, the waveguide is replaced by an equivalent volume current density which is nonzero only at locations which used to be occupied by the waveguide, and is zero elsewhere. From the Maxwell's equations (1a)-(1d), second order equations for electric and magnetic fields can be derived as
\begin{align}\label{eq:second order equations_1}
&\nabla^2\mathbf{E}+k^2\mathbf{E}=\frac {j}{\omega\epsilon_0\epsilon_s}(k^2\mathbf{J}+\nabla(\nabla\cdot\mathbf{J}))\\
&\nabla^2\mathbf{H}+k^2\mathbf{H}=-\nabla\times\mathbf{J},
\label{eq:second order equations_2}
\end{align}
where $k=\omega\sqrt{\epsilon_s\epsilon_0\mu_0}$ is the wave number in the surrounding medium. Equation~(\ref{eq:firstMaxwell}) can be written as
\begin{equation}\label{eq:firstMaxwell_2}
\nabla\times(\mathbf{E}\mathrm{e}^{j\gamma z}\mathrm{e}^{-j\gamma z})=-j\omega\mu_0\mathbf{H}.
\end{equation}
Using the identity for curl of product of a scalar and a vector, left hand side of (\ref{eq:firstMaxwell_2}) can be expended as
\begin{equation}\label{eq:firstMaxwell_expanded}
\nabla\times(\mathbf{E}\mathrm{e}^{j\gamma z})\mathrm{e}^{-j\gamma z}-j\gamma(\hat{z}\times\mathbf{E})=-j\omega\mu_0\mathbf{H}.
\end{equation}
Scalar multiplication of both sides of (\ref{eq:firstMaxwell_expanded}) by $\mathbf{H^*}$, and integrating over the waveguide cross section ($x-y$ plane) gives
\begin{align}\label{eq:S}
S=\frac{1}{2}\int\! \mathbf{E}\times\mathbf{H^*}\cdot\hat{z}\mathrm{d}a=\frac{\omega\mu_0}{2\gamma}\int\! |\mathbf{H}|^2\mathrm{d}a+ \frac{\mathrm{e}^{-j\gamma z}}{j2\gamma}\int\!\nabla \times (\mathbf{E}\mathrm{e}^{j\gamma z})\cdot\mathbf{H}^*\mathrm{d}a.
\end{align}
Here $\mathrm{d}a=\mathrm{d}x\mathrm{d}y$ is the differential of the cross sectional area, and the integrals in (\ref{eq:S}) and all other integrals in this paper are taken over the entire cross sectional plane. $S$ is the complex power passing through the waveguide cross section. It should be noted that for a mode with propagation constant of $\gamma$, $\mathbf{E}\mathrm{e}^{j\gamma z}$ has no $z$ dependence. It is useful to define
\begin{equation}\label{eq:F}
\mathbf{F}\triangleq\nabla\times(\mathbf{E}\mathrm{e}^{j\gamma z}).
\end{equation}
The material loss of the waveguide per unit length can be found using (\ref{eq:J})
\begin{align}\label{eq:P_l}
P_l=\frac{1}{2}\textrm{Re}\left\{\int\mathbf{E}\cdot\mathbf{J}^*\mathrm{d}a\right\}=
\frac{\epsilon_r
''}{2\omega\epsilon_0 |\epsilon_r-\epsilon_s|^2}\int|\mathbf{J}|^2\mathrm{d}a,
\end{align}
and using the Poynting's theorem for an infinitesimal slice of waveguide along its axis, the attenuation constant can be shown to be given by
\begin{equation}\label{eq:alpha}
\alpha=\frac{P_l}{2\textrm{Re}\{S\}}.
\end{equation}
Attenuation constant represents the waveguide propagation loss. We use the normalized second central moment of the magnetic energy density in the waveguide cross section as a criterion for the mode confinement. For a normalized mode with $\int|\mathbf{H}|^2\mathrm{d}a=1$, this central moment is defined as
\begin{equation}
\label{eq:sigma_H}
\sigma^2_\mathbf{H}=\int r^2|\mathbf{H}|^2\mathrm{d}a-(\int x|\mathbf{H}|^2\mathrm{d}a)^2-(\int y|\mathbf{H}|^2\mathrm{d}a)^2,
\end{equation}
where $r$ is radial distance from the $z$ axis of the coordinate system. We show that for a given value for the second central moment, a lower limit for the attenuation constant of the mode exists. To this end, we express the fields in the spectral domain.

\section{Spectral domain expressions for the waveguide loss and mode size}
\label{sec:Spectral domain formulation}
The spectral domain representation of a vectorial quantity such as $\mathbf{U}$ is defined as the Fourier transform of that vector in cross sectional plane of the waveguide
\begin{equation}
\tilde{\mathbf{U}}(k_x,k_y,z)=\int\mathbf{U}(x,y,z)\mathrm{e}^{-j(k_xx+k_yy)}\mathrm{d}a,
\end{equation}
and the inverse transform is given by
\begin{equation}
\label{eq:inverse fourier}
\mathbf{U}(x,y,z)=\frac{1}{4\pi^2}\int\tilde{\mathbf{U}}(k_x,k_y,z)
\mathrm{e}^{j(k_xx+k_yy)}\mathrm{d}s.
\end{equation}
In (\ref{eq:inverse fourier}), $\mathrm{d}s=\mathrm{d}k_x\mathrm{d}k_y$ represents the differential of area in the spectral domain and the integrals is over the entire $k_x-k_y$ plane. In the spectral domain, (\ref{eq:second order equations_1}) and (\ref{eq:second order equations_2}) can be written as
\begin{align}
&\tilde{\mathbf{E}}=\left[\begin{array}{c} \tilde{E_x}\\ \tilde{E_y}\\ \tilde{E_z} \end{array}\right]=\frac{1}{j\omega\epsilon_0\epsilon_s(k_r^2+\gamma^2-k^2)}
[\mathcal{M}]\tilde{\mathbf{J}}, \label{eq:E_tilde}\\
&\tilde{\mathbf{H}}=\left[\begin{array}{c} \tilde{H_x}\\ \tilde{H_y}\\ \tilde{H_z} \end{array}\right]=\frac{1}{(k_r^2+\gamma^2-k^2)}[\mathcal{B}]\tilde{\mathbf{J}}, \label{eq:H_tilde}
\end{align}
where $k_r^2=k_x^2+k_y^2$, and $[\mathcal{M}]$ and $[\mathcal{B}]$ matrices are given by
\begin{align}
[\mathcal{M}]&=\left(\begin{array}{ccc} k^2-k_x^2 & -k_xk_y & -k_x\gamma \\-k_xk_y & k^2-k_y^2 & -k_y\gamma  \\ -k_x\gamma  & -k_y\gamma  & k^2-\gamma^2 \end{array}\right), \\
[\mathcal{B}]&=\left(\begin{array}{ccc} 0 & j\gamma  & -jk_y \\-j\gamma &0 & jk_x \\ jk_y & -jk_x & 0 \end{array}\right).
\end{align}
Using (\ref{eq:F}) and (\ref{eq:E_tilde}) the spectral representation of $\mathbf{F}$ can be found as
\begin{equation}\label{eq:F_tilde}
\tilde{\mathbf{F}}=\frac{\mathrm{e}^{j\gamma z}}{j\omega\epsilon_0\epsilon_s(k_r^2+\gamma^2-k^2)}[\mathcal{N}]
\tilde{\mathbf{J}},
\end{equation}
where $[\mathcal{N]}$ is defined as
\begin{align}
[\mathcal{N}]=j\left(\begin{array}{ccc} k_xk_y\gamma & k_y^2\gamma & -k_y(k^2-\gamma^2) \\-k_x^2\gamma & -k_xk_y\gamma & k_x(k^2-\gamma^2)  \\ k^2k_y  & -k^2k_x  & 0 \end{array}\right).
\end{align}
Now, by use of the Parseval's theorem,
\begin{equation}\label{eq:power_theorem}
\int  \mathbf{V}^*\cdot\mathbf{U}\mathrm{d}a=\frac{1}{4\pi^2}\int\tilde{\mathbf{V}}^
\dagger\tilde{\mathbf{U}}\mathrm{d}s,
\end{equation}
$S$ and $P_l$ can be expressed in terms of spectral domain vectors. Multiplying each side of (\ref{eq:H_tilde}) by their Hermitian transpose gives
\begin{equation}\label{eq:H2}
|\tilde{\mathbf{H}}|^2=\tilde{\mathbf{H}}^\dagger\tilde{\mathbf{H}}=\frac{1}
{|k_r^2+\gamma^2-k^2|^2}\tilde{\mathbf{J}}^\dagger[\mathcal{B}]^
\dagger[\mathcal{B}]\tilde{\mathbf{J}}.
\end{equation}
By defining
\begin{equation}
[\mathcal{A}]\triangleq[\mathcal{B}]^\dagger[\mathcal{B}]
\end{equation}
and using (\ref{eq:power_theorem}), it can be found that
\begin{equation}\label{eq:int_H2}
\int|\mathbf{H}|^2\mathrm{d}a=\frac{1}{4\pi^2}\int\frac{1}{|k_r^2+\gamma^2-k^2|^2}
\tilde{\mathbf{J}}^\dagger[\mathcal{A}]\tilde{\mathbf{J}}\mathrm{d}s.
\end{equation}
From (\ref{eq:H_tilde}), (\ref{eq:F_tilde}), and (\ref{eq:power_theorem}) give
\begin{align}\label{eq:int_FH}
\int\!\mathbf{H}^*\cdot\mathbf{F}\mathrm{d}a=\!\!\frac{1}{4\pi^2}\!\!\int\!\tilde{\mathbf{H}}^
\dagger\tilde{\mathbf{F}}\mathrm{d}s
=\!\frac{-j\mathrm{e}^{j\gamma z}}{4\pi^2\omega\epsilon_0\epsilon_s}\!\int\!\frac{1}{|k_r^2+\gamma^2-k^2|^2}
\tilde{\mathbf{J}}^\dagger[\mathcal{B}]^\dagger[\mathcal{N}]\tilde{\mathbf{J}}\mathrm{d}s.
\end{align}
Substituting two left hand side integrals in (\ref{eq:int_H2}) and (\ref{eq:int_FH}) into (\ref{eq:S}) leads to
\begin{equation}\label{eq:S_spectral}
S=\frac{1}{8\pi^2\omega\epsilon_0\epsilon_s}\int\frac{1}{|k_r^2+\gamma^2-k^2|^2}
\tilde{\mathbf{J}}^\dagger[\mathcal{Z}]\tilde{\mathbf{J}}\mathrm{d}s,
\end{equation}
where $[\mathcal{Z}]$ is defined as
\begin{equation}
\label{eq:Z_definition}
[\mathcal{Z}]\triangleq\frac{1}{\gamma}(k^2[\mathcal{A}]-[\mathcal{B}]^
\dagger[\mathcal{N}]).
\end{equation}
Material loss of the waveguide can also be expressed in terms of
spectral domain volume current density. From (\ref{eq:P_l})
and (\ref{eq:power_theorem}), $P_l$ can be expressed in terms of spectral representation of the current density
\begin{equation}
P_l=\frac{\epsilon_r''}{8\pi^2\omega\epsilon_0|\epsilon_r-\epsilon_s|^2}\int |\tilde{\mathbf{J}}|^2\mathrm{d}s.
\end{equation}
Finally, the attenuation constant of the waveguide can be found
by substituting $S$ and $P_l$ from (\ref{eq:S_spectral})
and (\ref{eq:P_l}) into (\ref{eq:alpha})
\begin{equation}\label{eq:alpha_1}
\alpha=\frac{\epsilon_s\epsilon_r''}{2|\epsilon_r-\epsilon_s|^2}\frac{\int|
\tilde{\mathbf{J}}|^2\mathrm{d}s}{\int\frac{1}{|k_r^2+\gamma^2-k^2|^2}\textrm{Re}
\{\tilde{\mathbf{J}}^\dagger[\mathcal{Z}]\tilde{\mathbf{J}}\}\mathrm{d}s}.
\end{equation}

As a criterion for the mode size in the spectral domain, the normalized second central moment of the spectral domain magnetic field modulus squared can be used.
This moment is later related to the magnetic density moment defined in (\ref{eq:sigma_H}). For a normalized mode such that
\begin{equation}\label{eq:H2_normalization}
\int|\tilde{\mathbf{H}}|^2\mathrm{d}s=1,
\end{equation}
the second central moment of the magnetic field modulus squared in spectral
domain is defined similar to (\ref{eq:sigma_H}) as
\begin{equation}\label{eq:sigma_tilde}
\sigma^2_{\tilde{\mathbf{H}}}=\int k_r^2|\tilde{\mathbf{H}}|^2\mathrm{d}s-\left(\int k_x|\tilde{\mathbf{H}}|^2\mathrm{d}s\right)^2-\left(\int k_y|\tilde{\mathbf{H}}|^2\mathrm{d}s\right)^2.
\end{equation}

\section{An Upper limit on the waveguide propagation length}
\label{sec:Upper limit}
Here, we first find a lower bound for the attenuation constant for a given value of $\sigma^2_{\tilde{\mathbf{H}}}$, and then relate $\sigma^2_{\tilde{\mathbf{H}}}$  to $\sigma^2_{\mathbf{H}}$. To this end, we use eigenvectors and eigenvalues of $[\mathcal{A}]$  to choose $\tilde{\mathbf{J}}$ in a way to minimize the
attenuation constant. Eigenvalues and normalized eigenvectors of
$[\mathcal{A}]$ are found as
\begin{subequations}\label{eq:eig_A}
\begin{align}
\hat{V}_{\mathcal{A}_1}&=\frac{1}{k_r}\left[\begin{array}{c} k_y\\-k_x \\ 0 \end{array}\right], &\lambda&_{\mathcal{A}_1}=k_r^2+|\gamma|^2 \\
\hat{V}_{\mathcal{A}_2}&=\frac{|\gamma|}{k_r\sqrt{k_r^2+|\gamma|^2}}
\left[\begin{array}{c} k_x\\k_y \\-\frac{k_r^2}{\gamma^*} \end{array}\right], &\lambda&_{\mathcal{A}_2}=k_r^2+|\gamma|^2 \\
\hat{V}_{\mathcal{A}_3}&=\frac{1}{\sqrt{k_r^2+|\gamma|^2}}\left[\begin{array}{c} k_x\\k_y \\ \gamma \end{array}\right], &\lambda&_{\mathcal{A}_3}=0.
\end{align}
\end{subequations}
$[\mathcal{A}]$ is Hermitian and its normalized eigenvectors are three independent, mutually orthogonal vectors and constitute a orthonormal basis. Therefore, the volume current density in the spectral domain can be expanded in terms of them
\begin{equation}\label{eq:J_expansion}
\tilde{\mathbf{J}}=c_1\hat{V}_{\mathcal{A}_1}+c_2\hat{V}_{\mathcal{A}_2}+c_3
\hat{V}_{\mathcal{A}_3},
\end{equation}
where coefficients $c_i$ are functions of $k_x$ and $k_y$. Using (\ref{eq:alpha_1}) the attenuation constant can be expressed in terms of theses coefficients. More specifically, $|\tilde{\mathbf{J}}|^2$ in the numerator of the right hand side of this equation can be written as
\begin{equation}\label{eq:norm_J}
|\tilde{\mathbf{J}}|^2=|c_1|^2+|c_2|^2+|c_3|^2,
\end{equation}
and for expressing $\textrm{Re}\{\tilde{\mathbf{J}}^\dagger[\mathcal{Z}]\tilde{\mathbf{J}}\}$ in the denominator, the following relations can be used 
\begin{subequations}\label{eq:Zeig_A}
\begin{align}
[\mathcal{Z}]\hat{V}_{\mathcal{A}_1}&=(k^2\beta+j\alpha k^2)\hat{V}_{\mathcal{A}_1},\\
[\mathcal{Z}]\hat{V}_{\mathcal{A}_2}&=\left(k^2\beta+j\alpha( k^2-2k_r^2)\right)\hat{V}_{\mathcal{A}_2},\\
[\mathcal{Z}]\hat{V}_{\mathcal{A}_3}&=\frac{\gamma^*}
{|\gamma|}k_r(k^2-k_r^2-\gamma^2)\hat{V}_{\mathcal{A}_2}.
\end{align}
\end{subequations}
Equations (\ref{eq:Zeig_A}a)-(\ref{eq:Zeig_A}c) can be verified by direct substitution. Using these relations and (\ref{eq:J_expansion}), $\textrm{Re}\{\tilde{\mathbf{J}}^\dagger[\mathcal{Z}]\tilde{\mathbf{J}}\}$ can be found as
\begin{align}\label{eq:parabolic}
\textrm{Re}\{\tilde{\mathbf{J}}^\dagger[\mathcal{Z}]\tilde{\mathbf{J}}\}=
k^2\beta(|c_1|^2+|c_2|^2)+\textrm{Re}\{\frac{\gamma^*}{|\gamma|}k_r(k^2-k_r^2-\gamma^2)c_2^*c_3\}.
\end{align}
plugging $|\tilde{\mathbf{J}}|^2$ and $\textrm{Re}\{\tilde{\mathbf{J}}^\dagger[\mathcal{Z}]\tilde{\mathbf{J}}\}$ from (\ref{eq:norm_J}) and~(\ref{eq:parabolic}) into (\ref{eq:alpha_1}), results in   
\begin{align}\label{eq:alpha_2}
\alpha=\frac{\epsilon_s\epsilon_r''}{2|\epsilon_r-\epsilon_s|^2}\frac{\int(|c_1|^2+|c_2|^2)\mathrm{d}s+\int|c_3|^2\mathrm{d}s}{k^2\beta
\int\frac{|c_1|^2+|c_2|^2}{|k_r^2+\gamma^2-k^2|^2}\mathrm{d}s+\int\frac{\textrm{Re}
\{\frac{\gamma^*}{|\gamma|}k_r(k^2-k_r^2-\gamma^2)c_2^*c_3\}}
{|k_r^2+\gamma^2-k^2|^2}\mathrm{d}s}.
\end{align}

The square modulus of the spectral domain magnetic field can also be expressed in terms of the $c_i$ coefficients. Equations (\ref{eq:H2}) and (\ref{eq:J_expansion}) lead to
\begin{equation}\label{eq:H2_coefficient}
|\tilde{\mathbf{H}}|^2=\frac{k_r^2+|\gamma|^2}
{|k_r^2+\gamma^2-k^2|^2}(|c_1|^2+|c_2|^2).
\end{equation}
It can be observed from (\ref{eq:H2_coefficient}) that $|\tilde{\mathbf{H}}|^2$ and therefore $\sigma^2_{\tilde{\mathbf{H}}}$ does not depend on $c_3$. Thus, $c_3$ can be chosen freely to minimize the attenuation constant. In the (\ref{eq:alpha_2}), for a given value of $\int|c_3|^2\mathrm{d}s$ in the numerator, based on the Cauchy-Schwarz inequality, the integral involving $c_3$ in the denominator is maximized when
\begin{equation}\label{eq:c_3}
c_3=a\frac{k_r\gamma}{|\gamma|(k^2-\gamma^2-k_r^2)}c_2,
\end{equation}
where $a$ is a positive real number. Plugging in $c_3$ from (\ref{eq:c_3}) into (\ref{eq:alpha_2}) and expressing $|c_1|^2+|c_2|^2$ using (\ref{eq:H2_coefficient}) gives
\begin{equation}\label{eq:alpha_3}
\alpha\geq\frac{\epsilon_s\epsilon_r''}{2|\epsilon_r-\epsilon_s|^2}
\frac{\int\frac{|k_r^2+\gamma^2-k^2|^2}{k_r^2+|\gamma|^2}|
\tilde{\mathbf{H}}|^2\mathrm{d}s+a^2\int\frac{k_r^2|c_2|^2}{|k_r^2+\gamma^2-k^2|^2}\mathrm{d}s}
{k^2\beta\int\frac{1}{k_r^2+|\gamma|^2}|\tilde{\mathbf{H}}|^2\mathrm{d}s+
a\int\frac{k_r^2|c_2|^2}{|k_r^2+\gamma^2-k^2|^2}\mathrm{d}s}.
\end{equation}

Let us define:
\begin{subequations}
\begin{align}
b&\triangleq\int\frac{|k_r^2+\gamma^2-k^2|^2}{k_r^2+|\gamma|^2}|\tilde{\mathbf{H}}|^2\mathrm{d}s\\
u&\triangleq\int\frac{k_r^2|c_2|^2}{|k_r^2+\gamma^2-k^2|^2}\mathrm{d}s\\
d&\triangleq k^2\beta\int\frac{1}{k_r^2+|\gamma|^2}|\tilde{\mathbf{H}}|^2\mathrm{d}s
\end{align}
\end{subequations}
Using these definitions, (\ref{eq:alpha_3}) can be rewritten as
\begin{equation}\label{eq:alpha_3.5}
\alpha\geq\frac{\epsilon_s\epsilon_r''}{2|\epsilon_r-\epsilon_s|^2}
\frac{b+ua^2}{d+ua}.
\end{equation}
Right hand side of (\ref{eq:alpha_3.5}) is larger than its minimum value for different values of $a$ and this results in
\begin{equation}\label{eq:alpha_3.65}
\alpha\geq\frac{\epsilon_s\epsilon_r''}{2|\epsilon_r-\epsilon_s|^2}
\frac{b+ua^2}{d+ua}\geq\frac{\epsilon_s\epsilon_r''}{|\epsilon_r-\epsilon_s|^2}
\frac{1}{u}\left(\sqrt{bu+d^2}-d\right).
\end{equation}
It can be verified that $\frac{1}{u}\left(\sqrt{bu+d^2}-d\right)$ is a decreasing function of $u$ and its substitution by $q$ defined as
\begin{equation}
q\triangleq\int\frac{k_r^2(|c_1|^2+|c_2|^2)}{|k_r^2+\gamma^2-k^2|^2}\mathrm{d}s=
\int\frac{k_r^2}{k_r^2+|\gamma|^2}|\tilde{\mathbf{H}}|^2\mathrm{d}s\geq u, 
\end{equation}
results in a lower limit for $\frac{1}{u}\left(\sqrt{bu+d^2}-d\right)$ and therefore for $\alpha$, that is
\begin{equation}\label{eq:alpha_3.75}
\alpha\geq\frac{\epsilon_s\epsilon_r''}{|\epsilon_r-\epsilon_s|^2}
\frac{1}{q}\left(\sqrt{bq+d^2}-d\right).
\end{equation}

In many cases of interest $\frac{\alpha}{k}\ll 1$ and $\alpha^2$ can be neglected. Neglecting terms involving $\alpha^2$ in $b$, $q$, and $d$, defining $p$ as $p\triangleq\int k_r^2|\tilde{\mathbf{H}}|^2\mathrm{d}s+\beta^2-2k^2$, and using (\ref{eq:H2_normalization}), it can be found that
\begin{subequations}\label{eq:b_d}
\begin{align}
b&=p+\frac{k^4}{\beta^2}(1-q),\\
d&=\frac{k^2}{\beta}(1-q).
\end{align}
\end{subequations}
Plugging in $b$ and $d$ from (\ref{eq:b_d}) into (\ref{eq:alpha_3.75}) gives   
\begin{equation}\label{eq:alpha_3.95}
\alpha\geq\frac{\epsilon_s\epsilon_r''}{|\epsilon_r-\epsilon_s|^2}\frac{1}{q}
\left(\sqrt{pq+\frac{k^4}{\beta^2}(1-q)}-\frac{k^2}{\beta}(1-q)\right),
\end{equation}
and it can be readily verified that the right hand side of the inequality (\ref{eq:alpha_3.95}) is a decreasing function of $q$. It can also be observed from the definition of $q$ that $0<q\leq\int|\tilde{\mathbf{H}}|^2\mathrm{d}s=1$, therefore putting $q=1$ gives a lower bound for that expression, that is 
\begin{equation}\label{eq:alpha_3.97}
\alpha>\frac{\epsilon_s\epsilon_r''}{|\epsilon_r-\epsilon_s|^2}\sqrt{p}=
\frac{\epsilon_s\epsilon_r''}{|\epsilon_r-\epsilon_s|^2}\sqrt{\int k_r^2|\tilde{\mathbf{H}}|^2\mathrm{d}s+\beta^2-2k^2},
\end{equation}
and from (\ref{eq:sigma_tilde}) it is obvious that$\int k_r^2|\tilde{\mathbf{H}}|^2\mathrm{d}s\geq\sigma^2_{\tilde{\mathbf{H}}}$, therefore
\begin{align}\label{eq:alpha_8}
\alpha>\frac{\epsilon_s\epsilon_r''}{|\epsilon_r-\epsilon_s|^2}
\sqrt{\sigma^2_{\tilde{\mathbf{H}}}+\beta^2-2k^2}.
\end{align}

The uncertainty relation for the two dimensional Fourier transform requires
\begin{equation}\label{eq:uncertainty}
\sigma^2_{\mathbf{H}}\sigma^2_{\tilde{\mathbf{H}}}\geq 1
\end{equation}
combining (\ref{eq:alpha_8}) and~(\ref{eq:uncertainty}) and noticing that for guided modes $\beta>k$ (\ref{eq:alpha_8}) can be further simplified as
\begin{equation}\label{eq:alpha_5}
\alpha>\frac{\epsilon_s\epsilon_r''}{|\epsilon_r-\epsilon_s|^2}
\sqrt{\frac{1}{\sigma^2_{\mathbf{H}}}-k^2}.
\end{equation}
Expressing attenuation constant in terms of propagation length
($L=\frac{1}{\alpha}$), (\ref{eq:alpha_5}) becomes
\begin{equation}\label{eq:alpha_7}
\frac{L}{\lambda}<\frac{|\epsilon_r-\epsilon_s|^2}{\epsilon_s\epsilon_r''}
\frac{1}{\sqrt{\left(\frac{\lambda}{\sigma_{\mathbf{H}}}\right)^2-4\pi^2}}.
\end{equation}

\section{Discussion of the result and numerical examples}
\label{sec:discussion}

The inequality (\ref{eq:alpha_7}) shows that for a given waveguide mode size and a  waveguide material, there is an upper limit on the propagation length. As it was mentioned earlier, the limit is the result of the absorption loss of the waveguide  material. For achieving sub-diffraction limit confinement of waves, a nonzero equivalent current density should exist. As we showed, higher confinement requires a larger equivalent current density. Because the equivalent current density is supported by the waveguide core the larger equivalent current density means higher absorption loss; therefore, shorter propagation length.  

As it can be seen from (\ref{eq:alpha_7}), the material properties are only present as a multiplicative factor. In particular, for waveguides with air as surrounding material we can define a material loss merit factor as
\begin{equation}\label{eq:merit_factor}
M\triangleq\frac{|\epsilon_r-1|^2}{\epsilon_r''},
\end{equation}
which can be used for determining preferred waveguide materials. 

\begin{figure}[t]
\centerline{\includegraphics[width=0.8\columnwidth]{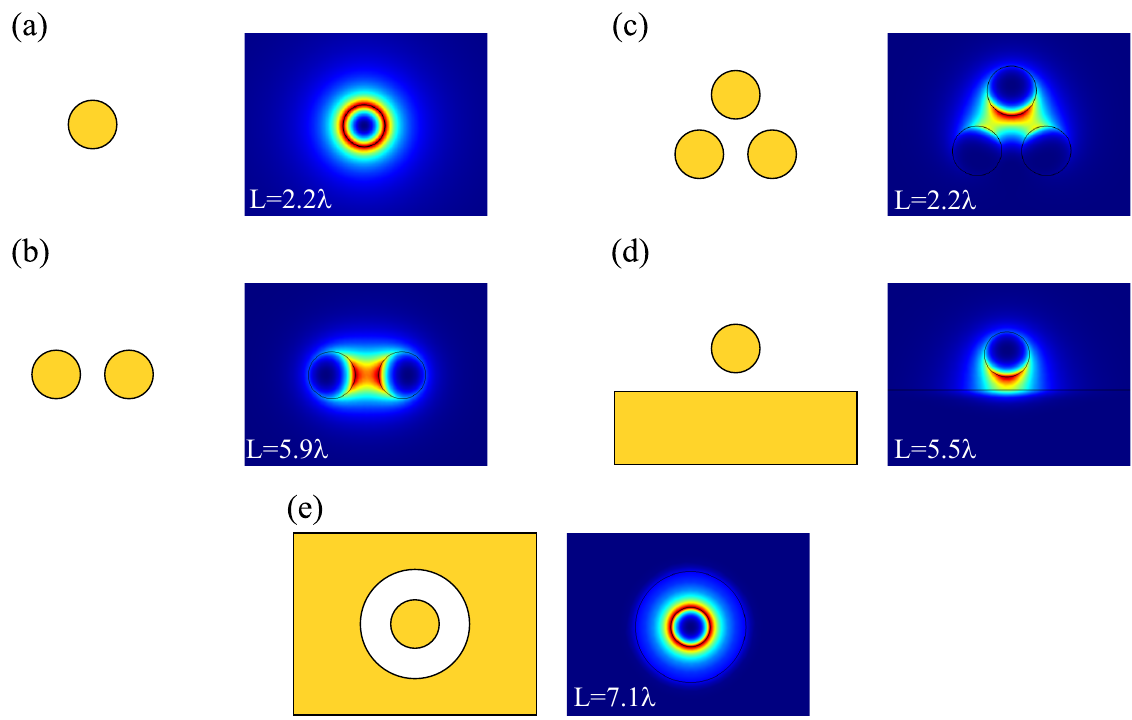}}
\label{fig:coaxial}
\caption{Schematic illustrations and simulated magnetic energy distributions for fundumental modes of five plasmonic waveguides. In the illustrations, the shaded areas are gold and the unshaded ones represent vacuum. The dimensions for all the wavegudies are chosen such that the mode sizes are all equal to 100 nm. The propagation lengths are shown at the lower corner of the energy distribution profiles.} \label{fig:waveguides}
\end{figure}

It is also interesting to compare the propagation lengths of a few plasmonic waveguides with the upper limit given in (\ref{eq:alpha_7}). According to (\ref{eq:alpha_7}), a waveguide made of gold and surrounded by vacuum with  subwavelength mode size of $\sigma_{{\mathbf{H}}}=100~\mathrm{nm}$ at  $\lambda=1.55~\mu\mathrm{m}$ ($\epsilon_{r_{Au}}\simeq-95.9-j11$ \cite{Palik1998}) has a propagation length shorter than $61.02\lambda$. Figure \ref{fig:waveguides} shows the schematics and magnetic energy distributions of few waveguide modes at $\lambda=1.55~\mu\mathrm{m}$. The waveguides are assumed to be made of gold and surrounded by vacuum. The magnetic energy density distribution and propagation constants of these modes are found using the Finite Element Method (FEM). All of the waveguides shown in the Fig. \ref{fig:waveguides} are designed to have mode size of $\sigma_{{\mathbf{H}}}=100~\mathrm{nm}$. The propagation lengths of the modes are also presented in the Fig. \ref{fig:waveguides}. As expected all of the propagation lengths are smaller than the theoretical upper limit. Coaxial waveguide (Fig. \ref{fig:coaxial}) has the longest propagation length among these waveguides. However, it is almost a factor of 9 smaller than the upper bound. More sophisticated waveguides and waveguides with graded index materials (with the same material loss merit factors) are expected to have longer propagation length and better achieve the upper bound.

\section*{Acknowledgment}
\label{sec:conclusion}
We would like to acknowledge supports from the National Science and
Engineering Council (NSERC) of Canada and Research In Motion (RIM).

\end{document}